\documentstyle[aps,prl,twocolumn,epsfig]{revtex}

\begin{document}

\draft

\title{Adiabatic compression of a trapped Fermi gas}

\author{L. Viverit$^{1}$, S. Giorgini$^{2}$, L.P. Pitaevskii$^{2,3}$ and S. Stringari$^{2}$}

\address{$^{1}$\O rsted Laboratory, H.C. \O rsted Institute, Universitetsparken 5, DK-2100
Copenhagen \O , Denmark}
\address{$^{2}$Dipartimento di Fisica, Universit\`a di Trento, \protect\\
and Istituto Nazionale di Fisica della Materia, I-38050 Povo, Italy}
\address{$^{3}$Kapitza Institute for Physical Problems, 117454 Moscow, Russia}


\maketitle

\begin{abstract}

{\it We propose a method to reach conditions of high degeneracy in a trapped Fermi gas,
based on the adiabatic transfer of atoms from a magnetic to a tighter optical trap. 
The transformation yields a large increase of the Fermi energy, without a significant change 
of the temperature. The large enhancement of the central density emphasizes the role of the 
interactions and makes the system much closer to the BCS transition. 
An estimate of the time needed to achieve the conditions of adiabaticity is also given.}

\end{abstract}


\narrowtext

The experimental realization of a highly degenerate atomic Fermi 
gas confined in traps is a  task of primary importance,  
especially in view of the perspective of approaching the
BCS transition to the superfluid phase. 
The regime of quantum degeneracy has been already reached
in a sample of potassium atoms \cite{Jila}, 
where first signatures of 
Fermi statistics, like the deviation of the velocity distribution from
a Boltzmann profile and the  increase of the kinetic energy with
respect to the classical value, have been observed.
The main difficulties in further lowering the temperature 
are due to the fact that the efficiency of the evaporative 
cooling process is strongly quenched \cite{Hol,Al}.  In fact Fermi statistics 
inhibits collisional processes at low temperature 
both directly, by reducing the phase space available for collisions, and
indirectly by lowering the density of the sample because of 
Pauli repulsion.
Procedures to optimize the evaporative process  
have recently permitted
to reach  lower temperatures, of the order of 
$0.2\sim0.3$ $T_F$ where $T_F$ is the Fermi temperature \cite{Hol}.

The purpose of this work is to propose a method to reach conditions of high degeneracy,
based on an  adiabatic compression of the gas. 
A similar method has proven quite successful in 
producing Bose-Einstein condensation in a reversible way, starting from
a trapped Bose gas above the critical temperature \cite{Mit}. The main point 
is that, by changing the shape of the confinement from a harmonic 
to a non harmonic trap, one can  increase  the degree of quantum degeneracy by
keeping the entropy of the total system constant \cite{Wal}.  
At the same time, the process of thermalization is not drastically quenched and
in typical experimental conditions takes place over times much shorter than the 
lifetime of the cloud.
In the following we 
consider a gas occupying two different spin states, 
initially confined by a harmonic trap. We then 
switch on adiabatically a second tighter trap (see Fig. 1). Experimentally
this can be realized using a magnetic trap for the
first confinement and an optical trap for the second one. As a
consequence of the adiabatic process a
fraction of atoms will move from the magnetic 
to the optical trap. This can  provide several  important advantages:

i) The gas in the  optical trap becomes
much more degenerate than the original one. In particular,
if the number of atoms transferred to the optical trap
is a small fraction,  the temperature 
will not change significantly with respect to the initial value, but the  
Fermi energy will increase,  
 in a way proportional to the depth of the optical trap.

ii) The  gas in the optical trap is much denser due to the tighter 
confinement. This produces an increase of interaction effects and
hence of the value of the critical temperature. 
Both 
effects i) and ii)  favour the reachability of
the BCS transition. 

iii) The Fermi energies of the two spin components, which initially
were different because of the different
magnetic trapping,
become closer in the optical trap, thereby favouring the
mechanism of Cooper pairing. 

Another important advantage of the proposed 
method
is that  the velocity distributions of the atomic clouds occupying the 
magnetic and optical traps can be measured separately. By releasing 
first the magnetic trap, one can measure the temperature 
of the sample. The effects of quantum degeneracy can then be 
investigated by measuring the velocity distribution
of the gas confined in  the optical trap.

In the first part of the work we assume that thermodynamic 
equilibrium is ensured during each step of the adiabatic compression
and we explore the 
properties of the new gas produced in the tight confinement by imposing 
entropy conservation. 
In the  second part of the work we   provide 
estimates for the times required to ensure adiabaticity and discuss
possible scenarios where  faster adiabatic trasformations 
take place out of thermal equilibrium.

Let us consider a gas initially confined in a harmonic trap
(hereafter called magnetic trap). We assume that the trapping 
frequencies and the number of atoms are the same for the two spin species. 
The Fermi energy is given by
$\epsilon_F^0 = \hbar\omega_{mag}(6N)^{1/3}$,
where $N$ is the number of atoms of each species and $\omega_{mag}$ is the  
geometrical average of the frequencies 
characterizing the magnetic trapping potential $V_{mag}$.
We will consider  systems lying initially in configurations of moderate 
degeneracy corresponding to $k_BT_i =0.2-0.5 \; \epsilon_F^0$, where $T_i$ is the 
initial temperature of the gas.
For simplicity, we will assume that also the optical 
trap can be approximated by a harmonic potential $V_{opt}$ having a tighter 
frequency $\omega_{opt}\gg\omega_{mag}$, and depth $V_{opt}({\bf r}=0) = - U$. 
In our model the trapping potential is then
defined as $V_{ext}({\bf r})=V_{opt}({\bf r})$ inside the optical trap 
($V_{opt}<V_{mag}$), and $V_{ext}({\bf r})=V_{mag}({\bf r})$ outside (see Fig. 1). 
The maximum number of atoms
that can be transferred in the optical trap is  given by the value
\begin{equation}
N_{opt} = \frac{1}{6}\left(\frac{U}{\hbar\omega_{opt}}\right)^3 \;\;.
\label{Nopt}
\end{equation}
As we will see, if we start from moderately low temperatures Eq.(\ref{Nopt})
provides an accurate estimate of the number of atoms which are 
actually transferred by the adiabatic process, provided $N_{opt}\ll N$.
The 
relative number of atoms transferred in the optical trap is then given 
by the useful expression
\begin{equation}
{N_{opt}\over N} = \left(\frac{U\omega_{mag}}{\epsilon_F^0\omega_{opt}}\right)^3 \;\;.
\label{Nrelat}
\end{equation}
Typical values that will be considered are  $\omega_{mag}/\omega_{opt} 
=0.1$ and $U/\epsilon_F^0 = 5$,
corresponding to $N_{opt}/N \sim $ 10\%.
Since the number of transferred atoms  is small one
expects that the final temperature $T_f$
of the gas will not change significantly with respect to the initial value $T_i$.
The final degree of degeneracy of the gas will be however significantly 
higher, 
since the final Fermi energy is approximately given by 
$\epsilon_F\simeq U+\epsilon_F^0$.
At the same time the central density of the gas, that at small temperature
is equal to $n(0)=(2m\epsilon_F^0/\hbar^2)^{3/2}/(6\pi^2)$ in the
initial stage, will increase by the factor $(\epsilon_F/\epsilon_F^0)^{3/2}$ 
in the optical trap. 
The increase of the Fermi energy and of the central  density 
has an important effect on the value of the BCS temperature
which is expected to behave, for negative scattering lengths,
as \cite{Stoof}
\begin{equation}
T_{BCS} \sim \frac{\epsilon_F}{k_B} \exp\left[-{\hbar\pi\over 2 p_F\mid a\mid }\right] \;\;,
\label{TBCS}
\end{equation}
where $p_F=\hbar(6\pi^2)^{1/3}n(0)^{1/3}$ is the 
Fermi momentum calculated in the center of the trap.
The value of the BCS temperature can increase significantly  
with respect to its value in the magnetic trap.
For example, by taking $a \sim -2000 a_0$, a ratio $U/\epsilon_F^0 =5$ and an initial
central density $n(0)\sim 10^{12}$ cm$^{-3}$, we obtain an increase of the 
density by a factor $\sim$15 and the BCS temperature (\ref{TBCS}) increases by the huge factor 
$\sim$50, becoming comparable to the initial value of the Fermi temperature $\epsilon_F^0/k_B$. 
The above discussion suggests that the proposed adiabatic mechanism
might provide conditions of high degeneracy, not far from the
transition to the BCS phase. 
With such a denser gas also the effects of the mean field on the 
density profile may be significant. An estimate is given by the ratio
\cite{VS} $E_{int}/E_{ho} \simeq 0.3 p_F a/\hbar$  
between the interaction energy and the oscillator energy of a spherically symmetric
trap. By using the values employed above 
one finds  corrections of the order of 10 \%.

The high degeneracy realized in the gas is expected to show up
in the
velocity distribution and in the release energy.
After completing the adiabatic transfer
one can release the magnetic trap. Measuring the velocity distribution
of these atoms then provides information on the temperature of the system
which is expected to be  close to the initial value. The atoms 
of the optical 
trap can be imaged in a second step.
For them one predicts that the ratio $E_{kin}/k_BT_f \sim 3U/8k_BT_f$ between the kinetic 
energy per particle and the thermal energy, 
should be  enhanced in a significant way if $U\gg k_BT_f$ revealing the effects of quantum 
degeneracy.
 
To confirm the scenario emerging from the above discussion we have carried
out a numerical calculation of the thermodynamic functions before
and after the adiabatic transformation. The calculation is obtained by 
imposing that the initial and final configurations have the same entropy.
This has been calculated using the semiclassical expression 
\begin{eqnarray}
\frac{S}{k_B} &=& \frac{1}{(2\pi\hbar)^3}\int d{\bf r}\, d{\bf p} \Big[ 
\frac{\epsilon({\bf p},{\bf r})/k_BT -\log z}{z^{-1}e^{\epsilon({\bf p},{\bf r})/k_BT}+1}
\Big. \nonumber \\
&+& \Big. \log\left(1+ze^{-\epsilon({\bf p},{\bf r})/k_BT}\right)\Big] \;\;,
\label{entropy}
\end{eqnarray}
where $z=\exp(\mu/k_BT)$ is the gas fugacity and
\begin{equation}
\epsilon({\bf p},{\bf r}) = \frac{p^2}{2m}+V_{ext}({\bf r})
\label{semiclen}
\end{equation}
are the semiclassical particle energies.
The results are presented in Figs. 2-5. 
In Fig. 2 we show
the relative number of atoms in the optical trap  as a function
of the depth of the optical trap $U/\epsilon_F^0$ for two initial temperatures. For the 
chosen configuration with $\omega_{mag}/\omega_{opt}=0.1$ and final depth 
$U = 5 \, \epsilon_F^0$, the optical trap  can host about 10 \% of atoms confirming the  
analytic prediction (\ref{Nrelat}).

In Fig. 3 the final temperature of the gas is plotted as a function of $U/\epsilon_F^0$.
As already anticipated the final temperature
of the gas does not change significantly from the original value, except for values 
of $U/\epsilon_F^0$ of the order of $\omega_{opt}/\omega_{mag}$. 
Notice however that, due to the tight confinement, the temperature of the gas can
become comparable to the optical oscillator temperature $\hbar\omega_{opt}/k_B$ if $N$
is not large enough. 

In Fig. 4 we show the kinetic energy per atom of the gas 
in units of the final temperature. The initial temperature is $k_BT_i=0.25 \epsilon_F^0$.
In the same figure, we also show the kinetic energy per particle of the gas occupying separately 
the  magnetic and the optical trap. These values are obtained 
by averaging separately the kinetic energy over the particles 
occupying the optical trap
($\epsilon({\bf p},{\bf r}) < 0$), and the particles occupying 
the magnetic trap ($\epsilon({\bf p},{\bf r}) > 0$).
The kinetic energy provides an important
indicator of the quantum degeneracy of the gas. The effect is very
spectacular for the atoms of the optical trap, where at $U=5 \epsilon_F^0$ one finds 
$E_{kin}/k_BT_f\simeq 7$. By choosing a higher initial temperature the effect is less 
pronounced, but still large ($E_{kin}/k_BT_f\simeq 3$ for $k_BT_i=0.5 \epsilon_F^0$ and 
$U=5 \epsilon_F^0$). 
Because of the high degeneracy 
the velocity distribution of the atoms in the optical trap deviates significantly 
from a Boltzmann distribution (Fig. 5).
We have also  calculated the central density as a function of the depth $U$. For 
$U=5 \,\epsilon_F^0$ and $k_BT_i=0.25 \epsilon_F^0$, we find an increase 
by a factor 15 with respect to the initial value.

In the second part of the work we discuss the conditions needed to
achieve the proposed adiabatic transformation. Let us recall
that there are several time scales in the problem. A first scale
is fixed by the periods of the harmonic wells. These times (of order of $10^{-3}$ - 
$10^{-1}$ sec) are expected to be shorter than the relaxation times due to 
collisions. The condition of reversibility requires that the relaxation time be much 
shorter than the time over which the optical trap is switched on. We have simulated the 
process of thermalization during the 
gradual increase of the depth of the optical trap by solving the quantum 
Boltzmann equation. We assume
equal distribution functions for the two spin components and that the phase-space 
distribution of particles is a function 
only of the single-particle energies $\epsilon({\bf p},{\bf r})$ (ergodic assumption) 
\cite{LRW96,Hol}.
This yields the following equation to solve
\begin{eqnarray}
\rho(\epsilon_1)\frac{\partial f(\epsilon_1)}{\partial t} &=& \frac{m\sigma}{\pi^2\hbar^3}
\int d\epsilon_3 d\epsilon_4 \;\rho(\epsilon_{min})
\nonumber \\ 
&\times& \left[ f(\epsilon_3)f(\epsilon_4)
(1-f(\epsilon_1))(1-f(\epsilon_2)) \right.\nonumber \\
&-& \left. f(\epsilon_1)f(\epsilon_2)(1-f(\epsilon_3))
(1-f(\epsilon_4))\right] \;\;,
\label{boltzmann}
\end{eqnarray}
where $\rho(\epsilon)$ is the density of states in the potential $V_{ext}$, which we model
by $\rho(\epsilon)=(\epsilon+U)^2/2(\hbar\omega_{opt})^3$ if $\epsilon<0$, and 
$\rho(\epsilon)=\epsilon^2/2(\hbar\omega_{mag})^3$ if $\epsilon>0$. Furthermore,
$\epsilon_{min} = $min$ \{\epsilon_1,\epsilon_2,\epsilon_3,\epsilon_4\}$ is the minimum value
of the four single-particle energies involved in the collisions  and $\epsilon_2=\epsilon_3
+\epsilon_4 -\epsilon_1$ according to energy conservation. The cross-section for collisions between 
the two distinguishable spin 
states is $\sigma=4\pi a^2$ and is fixed by the $s$-wave scattering length $a$.
In the numerical simulation the depth $U$ of the optical trap is ramped up by keeping the ratio 
$\omega_{mag}/\omega_{opt}$ constant, and at each step we let the gas thermalize in 
the new configuration. From Eq. (\ref{boltzmann}), it turns out that the time scale of thermalization
is fixed by $\tau^{-1}=\omega_{mag}(|a|/a_{mag})^2 N^{2/3}$, where $a_{mag}=
(\hbar/m\omega_{mag})^{1/2}$ 
is the magnetic oscillator length. If we choose $\omega_{mag}/\omega_{opt}=0.1$, 
$U=5\epsilon_F^0$ and 
$k_BT_i=0.5 \epsilon_F^0$, we find that the time required for the 
reversible transformation is $t_{rev}\simeq 100 \tau$. For a configuration  
with $N=10^6$, $\omega_{mag}=2\pi\times 100$ Hz and
$|a|/a_{mag}=5 \times 10^{-3}$, this corresponds to $t_{rev}\sim 1$ sec.    
If one starts from the lower initial temperature $k_BT_i=0.25 \epsilon_F^0$ and keeps the other
parameters unchanged, the  time $t_{rev}$ turns out to be about a factor two shorter.
The origin of this behavior is due to the mechanism of atom exchange between the magnetic
and the optical trap.  
Notice, however, that $t_{rev}$ depends rather strongly on the value of $N$ and becomes longer
if $N$ decreases.

The true shape of the optical trap differs from the harmonic potential employed above. 
Differences appear because a dipole trap can 
be safely approximated by a harmonic potential only at its center and, 
in general, can 
host more atoms than the corresponding harmonic trap with the same depth $U$ and frequency ratio 
$\omega_{mag}/\omega_{opt}$. This results in more heating and, as a consequence, in a higher final 
temperature $T_f$. For $k_BT_i=0.25 \epsilon_F^0$, we find that the fraction of atoms in the optical 
trap at $U=5\epsilon_F^0$ is about 50\% larger than the value obtained using the harmonic 
approximation, while the final temperature is only about 10\% higher. The final degree of degeneracy 
of the gas in the optical trap is consequently slightly reduced, but the qualitative features 
discussed above remain unchanged. 
   
If the time
of the transformation is longer than the inverse oscillator frequency, but faster
than the relaxation time (a relatively easy condition to realize experimentally),
then the transformation will be adiabatic, in the sense of entropy 
conservation, but the system will not be in thermal equilibrium. Such a
transformation corresponds to an adiabatic transfer of the lowest 
$N_{opt}$ single
particle states from the magnetic to the optical trap. The transformation keeps the 
corresponding occupation numbers unchanged. In this case, the atoms in the optical trap form  
a cold, out of equilibrium gas.
With the initial condition $k_B T_i=0.5 \, \epsilon_F^0$ 
this out-of-equilibrium transformation transfers approximately the same
fraction of atoms into the optical trap as with the fully reversible transformation discussed 
above and the average kinetic energy of these atoms takes a similar value, revealing that the 
system is highly 
degenerate. By using the quantum Boltzmann equation (\ref{boltzmann}) we have found that 
the time needed for this configuration to relax to equilibrium is significantly shorter, the 
final temperature of the gas being only sligthly higher than the one obtained in the reversible 
transformation.   

In conclusion we have investigated the consequences of an adiabatic transfer
of a gas of Fermions from a magnetic to a tighter optical trap. We have seen that it is
possible to reach configurations involving a significant fraction of
atoms which, as a result of the transformation, will occupy a 
Fermi sea in conditions of high degeneracy. The
large enhancement of the release energy should be easily observable
through time of flight measurements. 
This method could make it possible  to produce 
highly degenerate Fermi gases, which, in the case of gases interacting with 
negative scattering length, are close to the 
transition to the BCS phase.

Useful discussions with Brian DeMarco are acknowledged. It is also a pleasure to thank 
Wolfgang Ketterle for useful comments. L. V. wishes to thank the Danish Ministry of 
Education for support. This research was supported by Ministero dell'Universit\`a e della
Ricerca Scientifica e Tecnologica (MURST).

\begin{figure}
\begin{center}
\psfig{file=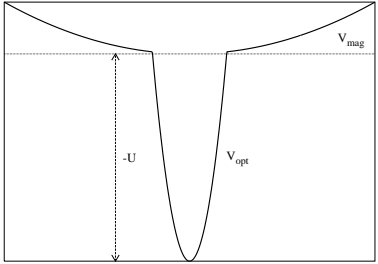,height=5.5cm,angle=-90}
\caption{Schematic representation of the confining potential $V_{ext}({\bf r})$ in the presence
of the magnetic and optical traps.}
\label{fig1}
\end{center}
\end{figure}

\begin{figure}
\begin{center}
\psfig{file=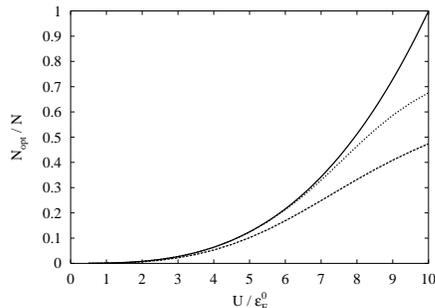,height=6.0cm,angle=-90}
\caption{Fraction of atoms in the optical trap as a function of $U/\epsilon_F^0$ for 
two initial temperatures: $k_BT_i=0.5\epsilon_F^0$ (dashed line), and $k_BT_i=0.25\epsilon_F^0$ 
(dotted line). The solid line corresponds to the analytical estimate (\ref{Nrelat}).}
\label{fig2}
\end{center}
\end{figure}

\begin{figure}
\begin{center}
\psfig{file=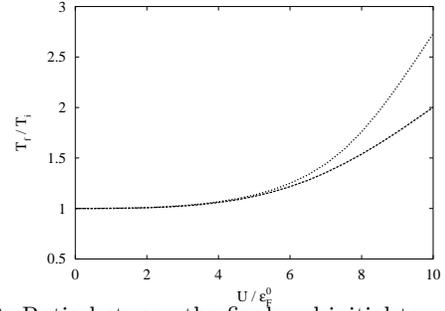,height=6.0cm,angle=-90}
\caption{Ratio between the final and initial temperature of the gas as a function of 
$U/\epsilon_F^0$ for the same initial temperatures as in Fig. 2.}
\label{fig3}
\end{center}
\end{figure}

\begin{figure}
\begin{center}
\psfig{file=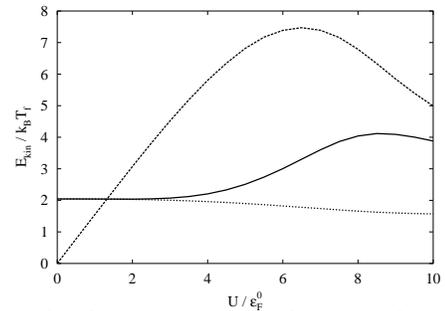,height=6.0cm,angle=-90}
\caption{Kinetic energy per particle $E_{kin}$ in units of the final temperature $k_BT_f$ as a 
function
of $U/\epsilon_F^0$. The initial temperature is $k_BT_i=0.25\epsilon_F^0$. The solid line 
corresponds to the kinetic energy per particle averaged over the entire system. 
Also shown are the kinetic energy per particle in the optical trap (dashed line), and in the 
magnetic trap (dotted line).}
\label{fig4}
\end{center}
\end{figure}

\begin{figure}
\begin{center}
\psfig{file=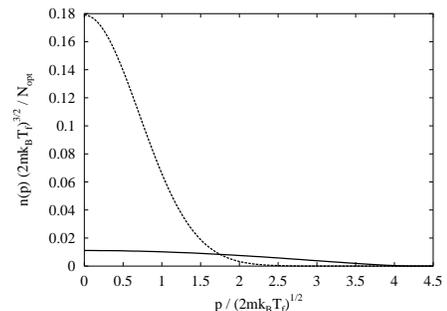,height=6.0cm,angle=-90}
\caption{Momentum distribution of the atoms occupying the optical trap with $U=5\epsilon_F^0$ 
(solid line). The initial value of the temperature is $k_BT_i=0.25\epsilon_F^0$. 
The dashed line corresponds to a Boltzmann gas of
the same number of atoms at the temperature $T=T_f$.}
\label{fig5}
\end{center}
\end{figure}    

\end{document}